\documentclass{article}
\usepackage{spconf,amsmath,graphicx,hyperref, amssymb, amsthm, cancel, bbding, multirow, fancyhdr}
\usepackage{dsfont}

\def\E{{\mathbb E}}
\def\R{{\mathbb R}}
\def\C{{\mathbb C}}

\def\Nr{{\cal N}}
\def\one{{\mathds{1}}}

\usepackage[table]{xcolor}
\usepackage[normalem]{ulem}
\usepackage{enumitem}

\newtheorem{theorem}{Proposition}

\title{Is phase really needed for weakly-supervised dereverberation ?}

\name{Marius RODRIGUES${}^*$ \qquad Louis BAHRMAN${}^*$ \qquad Roland BADEAU${}^*$ \qquad Gaël RICHARD${}^*$ \thanks{With support from the European Union (ERC, HI-Audio - Hybrid and Interpretable Deep neural
audio machines, 101052978).
Views and opinions expressed are however those of the author(s) only and do not necessarily reflect those
of the European Union or the European Research Council. Neither the European Union nor the granting
authority can be held responsible for them.}}
\address{${}^*$LTCI, Télécom Paris, Institut Polytechnique de Paris}

\fancypagestyle{firstpage}{

\fancyhead{}
\fancyfoot{}
\fancyfoot[C]{\footnotesize © 2026 IEEE. Personal use of this material is permitted. Permission from IEEE must be obtained for all other uses, in any current or future media, including reprinting/republishing this material for advertising or promotional purposes, creating new collective works, for resale or redistribution to servers or lists, or reuse of any copyrighted component of this work in other works.}
}

\begin{document}
\thispagestyle{firstpage}
\ninept
\maketitle
\begin{abstract} 
In unsupervised or weakly-supervised approaches for speech dereverberation, the target clean (dry) signals are considered to be unknown during training. In that context, evaluating to what extent information can be retrieved from the sole knowledge of reverberant (wet) speech becomes critical. 
This work investigates the role of the reverberant (wet) phase in the time–frequency domain. Based on Statistical Wave Field Theory, we show that late reverberation perturbs phase components with white, uniformly distributed noise, except at low frequencies.
Consequently, the wet phase carries limited useful information and is not essential for weakly supervised dereverberation. To validate this finding, we train dereverberation models under a recent weak supervision framework and demonstrate that performance can be significantly improved by excluding the reverberant phase from the loss function.
\end{abstract} 


\section{Introduction}
\label{sec:intro}

Acoustic waves propagating in enclosed environments are subject to \emph{reverberation}, a phenomenon arising from the repeated and decaying reflections of sound waves against room boundaries. In room acoustics, reverberation is commonly decomposed into three components: the \emph{direct path}, \emph{early reflections}, and \emph{late reverberation}.
Various models have been proposed to characterize reverberation \cite{välimäki2016more}, ranging from physics-based approaches~\cite{koutsouris2013combination}, \cite{polac1988},~\cite{badeau}, to perception-oriented frameworks \cite{schroeder}, \cite{jot1991digital}, or hybrid methods combining both perspectives \cite{bai2015fdnrtm}. More recently, the \emph{Statistical Wave Field Theory} has introduced a physics-based statistical formulation of late reverberation by analyzing the wave equation in the asymptotic regime of long times and high frequencies~\cite{badeau}, offering a principled framework particularly suited to the present study that focuses on late reverberation.

On the other hand, \textit{dereverberation} refers to the process of recovering a clean (dry) speech signal from its reverberant counterpart, and can be viewed as the inverse problem of reverberation. This task is of critical importance in several applications, including automatic speech recognition, hearing aids and teleconferencing systems where reverberation often degrades both intelligibility and performance.

As the degradation model is generally not known in practice, this ill-posed problem has been recently solved by modeling either the source-receiver system explicitly~\cite{aknin2021stochastic}, \cite{lalay_modephysique_2025}, \cite{belhomme} or only the source in implicit data-driven approaches~\cite{tflocoformer}, \cite{fullsubnet}, \cite{bahrman_ICASSP}.

In generic speech enhancement, it has been demonstrated that integrating the phase spectrum into the enhancement process can improve the resulting quality~\cite{importance-phase}, \cite{gerkmann2015phase-processing}.
For that reason, current state-of-the-art machine learning approaches for speech enhancement such as TF-GridNet~\cite{tf-gridnet} or FullSubNet~\cite{fullsubnet} include the phase in their training objective. However, these approaches rely on full supervision and require access to the dry phase during training~\cite{tf-gridnet}, \cite{tflocoformer}, \cite{schwartz-mag-or-phase}.
In scenarios of weak or self supervision, phase reconstruction is expected to be more challenging, since the algorithm can only rely on the knowledge of reverberant signals, in which phase has been corrupted.
Some recent approaches seem to circumvent this problem by discarding reverberant phase information~\cite{fullsubnet}, reuse the reverberant phase as dry phase~\cite{bahrman_ICASSP}, or using phase retrieval as a refining post-processing step~\cite{griffin-lim}, \cite{skerry-ryan2017tacotron}, \cite{li2021two-heads-are-better-than-one}, \cite{zhao2019two-stage-speech-enhancement}. Indeed, it was demonstrated that keeping the distorted phase may be the best option~\cite{ephraim1984speechMMSE}, under a circularity assumption of measured signals. 

In this paper, we demonstrate that this assumption physically holds for reverberation distortion. More precisely, we quantify to what extent reverberation degrades the phase of the dry signal, and how unsupervised dereverberation approaches should be adapted to these constraints. Our main result is that, in theory and in practice, reverberant signals, under reasonable acoustic conditions, hold almost no information about the dry phase. The contribution of this paper is then twofold:
\begin{itemize}
    \item We establish, based on the \emph{Statistical Wave Field Theory}, that phase perturbations induced by late reverberation are statistically uniform and white in the time–frequency domain. This finding arises from the physics of wave propagation and is not restricted to any specific signal-processing-based reverberation model.
    \item We experimentally demonstrate the benefit of incorporating phase-invariance into model training when access to clean reference signals is limited. This analysis is conducted within the weakly supervised framework introduced in \cite{bahrman_ICASSP}, which provides a representative case study for validating our claim.
\end{itemize}

These results suggest that weakly or self-supervised models may achieve significant gains simply by ignoring reverberant phase information. For reproducibility purposes and to help future research, detailed mathematical proofs, code, pretrained models and audio examples are made publicly available\footnote{https://mariusrod.github.io/PhaseInv-WSSD/} (code will be openly distributed upon acceptance).

\section{Statistical modeling}
\label{sec:stat_model}

From a signal processing point of view, the reverberation phenomenon can be considered as a filtering operation. Denoting as $s$ the original dry signal, and $h$ the so-called Room Impulse Response (RIR), the resulting reverberant signal is the convolution of $s$ with $h$:
\begin{equation}y=s\star h.\end{equation}

From a physical point of view, reverberation is the diffuse mixing of a sound wave with its echoes after many reflections against the room's walls. Recently introduced, the \emph{Statistical Wave Field Theory} proposes a probabilistic approach to solving the wave equation \cite{badeau}. Its results generalize those from \cite{polac1988}, allowing to derive the first and second moments of any RIR in long times and at high frequencies.

The broadly used simplified Polack model expresses the late RIR as an exponentially decaying Gaussian white noise:
\begin{equation}
  h(t) = \sqrt{B}\varepsilon(t)e^{-\frac{\alpha}2 t}, 
  \label{eq:polack}
\end{equation}
where $\varepsilon(t) \sim\mathcal N(0,1)$, $B$ is the late reverberation magnitude, and $\alpha$ is the decaying rate, related to the so-called reverberation time $RT_{60}$ through $\alpha = \frac{6\ln(10)}{RT_{60}}$. 
While this model is perceptually accurate, observations show that the parameters $\alpha$ and $B$ are actually frequency-dependent.
One of the key results of the Statistical Wave Field Theory confirms that fact and establish the physically-accurate \emph{generalized} Polack model. With formal words, a Polack RIR $h$ with frequency-dependent parameters $\alpha(f)$ and $B(f)$ can be described as follows:
\begin{itemize}
    \item $h$ is Gaussian and centered
    \item and its autocovariance $\gamma_h(t,\tau):= \E[h(t+\frac\tau2)h(t-\frac\tau2)]$ is given by: \begin{equation}\gamma_h(t,\tau)=\one_{\R^+}(t)\int B(f)e^{-\alpha(f)t+2i\pi f\tau}\text{d}f.\end{equation}
\end{itemize}
Note here that $B(f)$ and $\alpha(f)$ are necessarily symmetric to ensure $h$ is real.
Another interesting property is that any scalar product $\langle h, \varphi\rangle:= \int h\varphi$ is also Gaussian, with variance \begin{equation}\label{eq:variance}\sigma^2(\varphi) = \iint\varphi\left(t+\frac\tau2\right)\varphi\left(t-\frac\tau2\right)\gamma_h(t,\tau)\text d\tau\text dt.\end{equation}
In a more general way, the covariance between two scalar products is given by:
\begin{equation}\label{eq:covariance}\text{Cov}(\langle h, \varphi\rangle, \langle h, \psi\rangle) = \iint\varphi\left(t+\frac\tau2\right)\psi\left(t-\frac\tau2\right)\gamma_h(t,\tau)\text d\tau\text dt.\end{equation}

\section{Phase in reverberant signals}
\label{sec:methods}

\begin{figure*}[htb]
\centering
\begin{minipage}[b]{0.3\linewidth}
  \centering
  \centerline{\includegraphics[width=5.0cm]{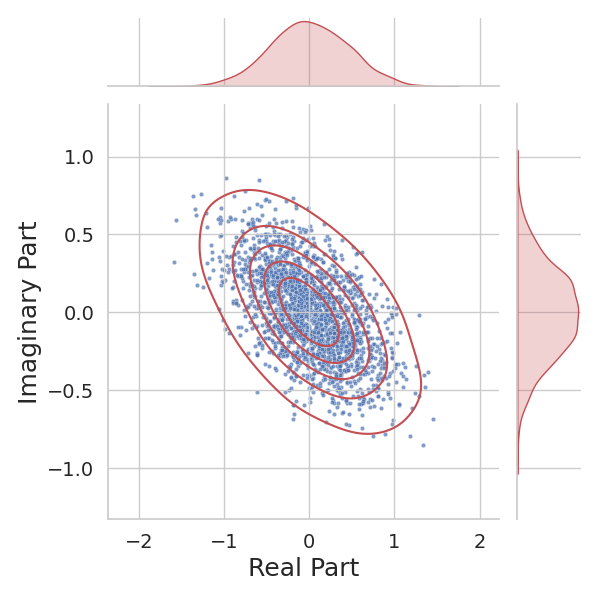}}
  \centerline{(a) $f=10 ~ Hz$}\medskip
\end{minipage}
\begin{minipage}[b]{.3\linewidth}
  \centering
  \centerline{\includegraphics[width=5.0cm]{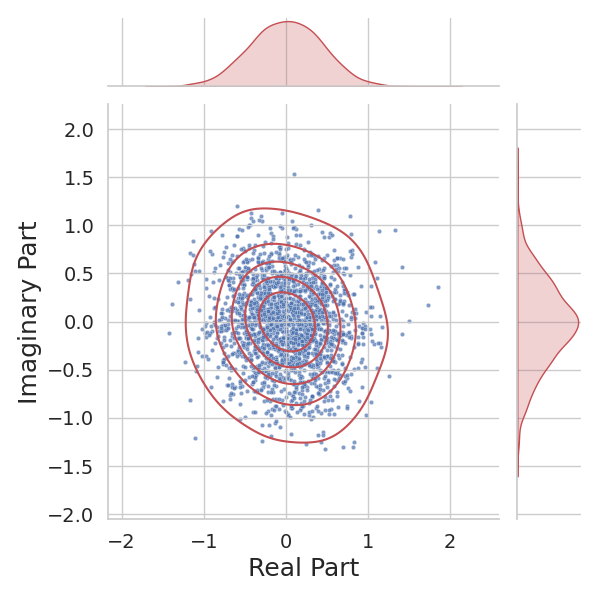}}
  \centerline{(b) $f=100 ~ Hz$}\medskip
\end{minipage}
\begin{minipage}[b]{0.3\linewidth}
  \centering
  \centerline{\includegraphics[width=5.0cm]{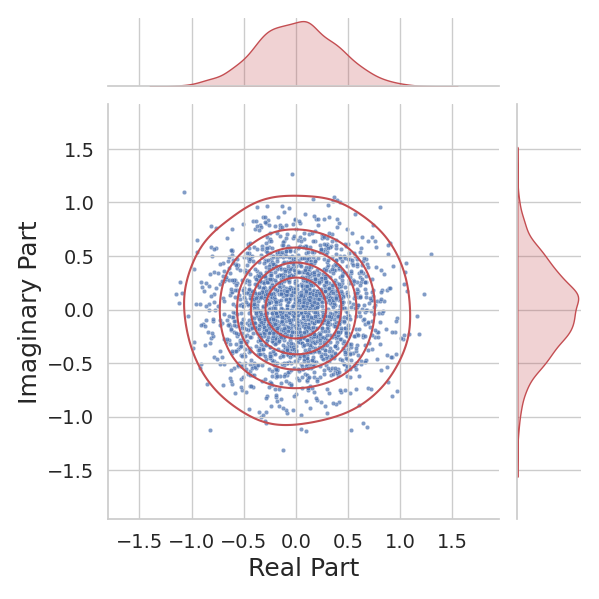}}
  \centerline{(c) $f=1000 ~ Hz$}\medskip
\end{minipage}

\caption{Distribution of the Fourier coefficients of a synthetic RIR in the complex plane, at 3 different frequencies. To parametrize the RIR, $\alpha(f)$ and $B(f)$ are autoregressive (AR) profiles of order 8 whose poles are randomly chosen in the unit disk. Here, their means over $f$ are respectively $\bar\alpha \simeq 0.0113 s^{-1}$ (corresponding to a reverberation time of approximately $82~ ms$) and $\bar B \simeq 0.0029$}
\label{fig:distribs}
\end{figure*}

In this section, we present the main result of our paper: late reverberation introduces a uniform phase perturbation in the reverberant signal.

\subsection{Asymptotical phase distribution in the Fourier space}\label{ssection:main_result}

\begin{theorem}\label{th:asymp_distrib}
In formal words, if we consider the RIR $h$ to be randomly sampled according to the generalized Polack model, as in Section \ref{sec:stat_model}, with parameters $\alpha(f)$ and $B(f)$, and $H$ its Fourier transform, then, asymptotically when $f\to+\infty$, 
        \begin{equation}\label{eq:distrib_asymp}H(f)\sim\Nr_\C\left(0,\frac{B(f)}{\alpha(f)}\right).\end{equation}
        
\end{theorem}

The proposition can be proven by expressing the Fourier transform of $h$ along the real and imaginary axes. Denoting $c_f(t) = \cos(2\pi ft)$ and $s_f(t) = \sin(2\pi ft)$, we get:
    \begin{align}
        H(f) &:= \int h(t)e^{-2i\pi f t}\text{d}t\\
        &=\langle h, c_f\rangle - i\langle h, s_f\rangle.
    \end{align}

Following the Statistical Wave Field Theory, $H(f)$ can be viewed as a bi-dimensional centered Gaussian random variable whose covariance matrix $\Sigma(f):= \begin{pmatrix}\sigma^2_{+}(f) & C(f)\\C(f)&\sigma_-^2(f)\end{pmatrix}$ can be computed thanks to the variance and covariance expressions from Section \ref{sec:stat_model}. With help of integration by parts and linearization formulas, one can end up with:
\begin{align} 
    \sigma_\pm^2(f) &=\frac{B(f)}{2\alpha(f)}\pm\frac{B(0)}{2\alpha(0)\left[1+\left(\frac{4\pi f}{\alpha(0)}\right)^2\right]}\\
    &=\frac{B(f)}{2\alpha(f)}\pm O\left(\frac{B(0)\alpha(0)}{f^2}\right),
\end{align}
and
\begin{align}
    C(f) =\text{Cov}(\langle h, c_f\rangle, \langle h, s_f\rangle) &= \frac{2\pi fB(0)}{\alpha(0)^2\left[1+\left(\frac{4\pi f}{\alpha(0)}\right)^2\right]}\\
    &=O\left(\frac{B(0)}{f}\right).
\end{align}
Hence, for $f\to +\infty$,
\begin{itemize}
    \item $\sigma_+^2(f) \simeq \sigma_-^2(f) \simeq \frac{B(f)}{2\alpha(f)},$
    \item $C(f)\simeq 0,$
\end{itemize}
which ends the proof.

Denoting $\angle z$ the argument of a complex number $z$, a direct consequence of Proposition~\ref{th:asymp_distrib} is that, asymptotically $\angle H(f)\sim\text{Unif}([0,2\pi])$. As the reverberant signal is $y = h\star s$, if $Y$ and $S$ are the Fourier transforms of $y$ and $s$, \begin{equation}\angle Y(f) = (\angle H(f) + \angle S(f)) ~\text{mod}~ 2\pi,\end{equation} which yields $\angle Y(f) | \angle S(f) \sim \text{Unif}([0,2\pi])$ for $f\to+\infty$.

\subsection{Visualization on synthetic data}
As the previous result is asymptotic, one could argue that the phenomenon occurs only at high frequencies. However, it should be noted from the previous demonstration that the convergence rates of the variances $\sigma_\pm^2(f)$ are quadratic. Figure \ref{fig:distribs} shows the statistical distribution of $H$ at specific frequencies $f$ ($10, 100$ and $1000 ~ Hz$). To sample the RIRs under parameters $\alpha$ and $B$, we modulated a centered Gaussian white noise $\varepsilon$ on separated frequency bands $[f_i, f_{i+1}]$:
\begin{equation}h[t] = \sum_{i=1}^{N}\sqrt{B(f_i)}(\varepsilon\star\varphi_i)[t]e^{-\frac{\alpha(f_i)}{2}t},\end{equation}
where $\{\varphi_i\}_i$ is a Butterworth filter bank on the frequency subdivision $0 < f_1 < ... < f_{N-1}<\frac{F_s}2$, $F_s$ being the sampling frequency. We observe that the probability distribution of $H(f)$ starts to be isotropic around $100 ~ Hz$ for a mean $RT_{60}$ as low as $82~ ms$. Longer reverberation times would lead to even faster convergence, with isotropy appearing at frequencies to the order of $10~Hz$.

\subsection{Extension to the time-frequency domain}
Taking inspiration from the proof presented in Section \ref{ssection:main_result}, an expression can be computed for the autocorrelation of the RIR in the Fourier space:
\begin{equation}\label{eq:correlations}
    \E\left[H\left(f+\frac{\xi}2\right)H^*\left(f-\frac{\xi}2\right)\right] = \frac{B(f)}{\alpha(f)}\cdot\frac{1-i\frac{2\pi\xi}{\alpha(f)}}{1+\left(\frac{2\pi\xi}{\alpha(f)}\right)^2}.
\end{equation}
Note that, with a similar convergence rate as before, when $\xi\to+\infty$, the correlation between frequency bins goes to~$0$. Similarly as before, reasonable values for $\alpha$ imply that a given frequency coefficient $H(f)$ is only correlated to its neighbors within a thin frequency band. Outside of this interval, the Fourier coefficients can be considered as independent. In practice, signals are analyzed in the spectro-temporal domain with a frequency bandwidth that is at the order of or greater than $10 ~ Hz$. A similar remark should apply along the time axis, as long as the window size is large enough. In the end, if $\angle H$ appears to be very close to a white uniform noise, this is even more the case for time-frequency representations such as the Short-Time Fourier Transform (STFT).

All-in-all, these theoretical results suggest that the reverberant's phase mostly carries unrelevant noise. An important remark to consider along this claim is that our results are derived from the study of a physics-based statistical description of late reverberation. Hence it is more than just a modeling bottleneck, and should apply to real-life RIRs.
In consequence, as long as no access is guaranteed neither on the original dry signal nor the exact phase perturbation, the reverberant's phase information should be used with caution or not at all.

\section{Experiments and discussion}
\label{sec:typestyle}

\begin{table*}[htb]\label{tab:metrics}
\centering
\begin{tabular}{c|c|cccc}
\hline
 Model         & $f_i(z)$   & SRMR $\uparrow$ & SISDR $\uparrow$ & WB-PESQ $\uparrow$ & ESTOI $\uparrow$ \\
\hline\hline
 \multirow{4}{4em}{\centering FSN} &  $z$ & 3.859 ± 1.76 & -16.719 ± 1.07e+02 & 1.291 ± 1.82e-02 & 0.572 ± 5.80e-03 \\
                        &  $\log(1+|z|)\frac{z}{|z|}$ & 3.246 ± 1.85 & -17.663 ± 7.94e+01 & 1.248 ± 1.51e-02 & 0.553 ± 5.94e-03 \\
                        &  $|z|$ & 6.024 ± 3.00 & \underline{-16.252 ± 1.12e+02} & 1.381 ± 2.63e-02 & 0.642 ± 4.33e-03 \\
                        &  $\log(1+|z|)$  & \underline{6.563 ± 3.88} & -16.541 ± 9.78e+01 & \underline{1.405 ± 3.08e-02} & \textbf{0.647 ± 4.41e-03} \\
 \hline
 PI-FSN            &   $\log(1+|z|)$  & \textbf{6.604 ± 3.20} & \textbf{-2.111 ± 2.95}  & \textbf{1.428 ± 3.56e-02} & \underline{0.645 ± 4.25e-03} \\
 \hline
    \rowcolor{lightgray}\multicolumn{2}{c|}{Input Reverberant}             & 4.357 ± 2.49 & -16.539 ± 1.05e+02 & 1.323 ± 2.38e-02 & 0.584 ± 5.77e-03 \\
\hline
\end{tabular}
\caption{Evaluation of the dereverberation models FSN and PI-FSN after training by weak-supervision, under several configurations of the loss. The equivalent evaluation of a model returning its input is given at the last row. For every metric, the higher the better, the best result is in bold form and second best is underlined.}
\end{table*}

This section aims to show how the previously presented reverberation-induced uniform phase perturbation can influence the training of dereverberation models.

\subsection{Global settings}
In our experiments, we chose to challenge the weakly-supervised speech dereverberation method recently introduced in \cite{bahrman_ICASSP}. The global pipeline of this method can be compared to the strategy of an auto-encoder. In place of the encoder, a neural network computes an estimation $\hat s$ of the dry signal, given the reverberant signal $y = s\star h$. In place of the decoder, a RIR synthesizer is given the reverberation parameters $\alpha$ and $B$ to produce a secondary RIR $\hat h$ that, when convolved to $\hat s$, produces an estimate $\hat y$ of the input reverberant signal. The dereverberation model is then trained by optimizing a reconstruction loss $\mathcal L(y,\hat y)$. 
Hence, it has no access to dry phase information during training, and can be considered a weakly-supervised approach in that regard.
For our purposes, we chose the following settings:

\begin{itemize}
    \item \textbf{Models:} the dereverberator is chosen to be \emph{FullSubNet} (FSN) \cite{fullsubnet}, an LSTM-based speech enhancement model used to test the above weakly supervised strategy in the original paper \cite{bahrman_ICASSP}. Its ability to combine full-band and sub-band spectro-temporal features makes it a good candidate to be paired with reverberation-informed training strategies. The original FSN takes as input the STFT magnitude of the reverberant signal and computes an estimation of the ideal \emph{complex Ratio Mask} (cRM) for retrieving the dry signal. For our purposes, we will also consider a phase-invariant version of FullSubNet (PI-FSN), by discarding the complex masking for a real positive one that conserves the original phase. For the models' hyperparameter configuration, we simply follow the one from~\cite{bahrman_ICASSP}.
    \item \textbf{RIR synthesizer:} as in \cite{bahrman_ICASSP}, an RIR is drawn using~\ref{eq:polack}, and depends on the decay rate $\alpha$ and magnitude $B$, which are supposedly known at training.
    \item \textbf{Losses:} since the main purpose of this study is to demonstrate the effect of phase in model training, the considered losses are of the form:
    \begin{equation}\mathcal L(\mathcal Y, \hat{\mathcal Y}) = \E[\Vert f_i(\mathcal Y) - f_i(\hat{\mathcal Y}) \Vert^2],\end{equation} where $\mathcal Y$, $\hat{\mathcal Y}$ are the STFTs of $y$ and $\hat y$, and $f_i$ is one of the following post-processing functions:
    \begin{itemize}[noitemsep, label=$\circ$]
         \item $f_1(z) = z$ \textit{(phase is kept + no compression)}
        \item $f_2(z)=\log(1+|z|)\frac{z}{|z|}$ \textit{(phase is kept + log-compress.)}
        \item $f_3(z)=|z|$ \textit{(phase discarded + no compression)}
        \item $f_4(z)=\log(1+|z|)$ \textit{(phase discarded + log-compress.)}
    \end{itemize}
\end{itemize}

\subsection{Training and testing data}
The dataset used for our experiments is EARS-Reverb \cite{richter2024ears}, a collection of more than 100 hours of English speech recordings with various intonations, intentions, and voicing styles. To this database is joint a dereverberation benchmark gathering several sets of real-life RIRs \cite{ARNI}\cite{ACE-Challenge}\cite{AIR}\cite{BRUDEX}\cite{dEchorate}, totaling more than 2000 recordings. Reverberant samples are obtained by randomly selecting dry/RIR pairs $(s_i, h_i)$ and then convolving $y_i = s_i\star h_i$. $15\%$ of the dataset is kept apart for testing and $8\%$ for validation during training. The validation and test sets are built with RIRs from separated sets of the EARS-Reverb benchmark to ensure the used room acoustics are totally new to the model. 
\subsection{Evaluation and results}

The performances of the trained models are evaluated using the Scale-Invariant Signal to Distortion Ratio (SISDR) for the global estimation error \cite{SISDR}, the Speech to Reverberant Modulation energy Ratio (SRMR) for reverberation-like distortion \cite{SRMR}, the Wide-Band Perceptual Estimation of Speech Quality (WB-PESQ) \cite{PESQ}, and the Extended Short-Time Objective Intelligibility (ESTOI)\cite{STOI}. The results are presented in Table \ref{tab:metrics}. All configurations of the loss function were tested with a simple FSN, while only the best configuration obtained for FSN (compression + phase invariance) was tested with PI-FSN.

First, looking at FSN's performances, the positive effect of introducing phase invariance in the loss is very clear. In every metric and regardless of the use of compression, best results are obtained when the information of phase is discarded. The difference is particularly significant in SRMR, indicating that a phase invariance loss is specifically helpful for dereverberation tasks. On the other hand, compression plays a rather little role, but seems to slightly help when combined with phase-invariance. Finally, the reader might pay a close attention to the score in SISDR for FSN, since its values are of the same order than the one of the input reverberant signal. Indeed, this version of FullSubNet is designed to estimate the ideal cRM, meaning that it might try to estimate the phase while it actually cannot. On the contrary, introducing phase invariance to the model's estimation (PI-FSN) yields a strong boost in SISDR, while keeping similar results in other metrics. All in all, those results are aligned with our main claim: the wet phase is so corrupted by late reverberation that it does not hold much useful information. Additionally, our results point to the fact that using the reverberant signal's phase might even deteriorate the model's performances.  

\section{Conclusion}
\label{sec:conclusion}

In this work, we explored the effects of using the phase information in weak or self-supervised learning strategies for dereverberation. Using results from the Statistical Wave Field Theory, we mathematically proved that the reverberation-induced phase perturbation is similar to a uniform white noise, which, to the best of our knowledge, has never been done before. Consequently, we experimentally observed in the context of weak supervision that models could significantly benefit from the introduction of phase invariance. Future works could extend this study to other weak or self-supervision scenarios, or design new approaches with our results in mind. For example, we could stack a phase-invariant dereverberator, estimating the dry spectrogram magnitude, with a phase retrieval model. While the dereverberator could be trained with weak supervision, the phase estimator could be pretrained on a much larger dataset.

\bibliographystyle{IEEEbib}
\bibliography{main}  

\begin{thebibliography}{10}

\bibitem{välimäki2016more}
Vesa Välimäki, Julian Parker, Lauri Savioja, Julius~O. Smith, and Jonathan
  Abel,
\newblock ``More than 50 years of artificial reverberation,''
\newblock {\em Journal of the audio engineering society}, , no. K-1, January
  2016.

\bibitem{koutsouris2013combination}
Georgios~I Koutsouris, Jonas Brunskog, Cheol-Ho Jeong, and Finn Jacobsen,
\newblock ``Combination of acoustical radiosity and the image source method,''
\newblock {\em J. Acoust. Soc. Am.}, vol. 133, no. 6, pp. 3963--3974, 2013.

\bibitem{polac1988}
Jean-Dominique Polack,
\newblock {\em La transmission de l'energie sonore dans les salles (in
  french)},
\newblock Ph.D. thesis, Universit{\'e} du Maine, 1988.

\bibitem{badeau}
Roland Badeau,
\newblock ``Statistical wave field theory,''
\newblock {\em J. Acoust. Soc. Am.}, vol. 156, no. 1, pp. 573--599, July 2024.

\bibitem{schroeder}
Manfred~R Schroeder and Ben Logan,
\newblock ``"colorless" artificial reverberation,''
\newblock {\em IRE Transactions on Audio}, vol. AU-9, no. 6, pp. 209--214,
  1961.

\bibitem{jot1991digital}
Jean-Marc Jot and Antoine Chaigne,
\newblock ``Digital delay networks for designing artificial reverberators,''
\newblock {\em Journal of the audio engineering society}, , no. 3030, February
  1991.

\bibitem{bai2015fdnrtm}
Hequn Bai, Gaël Richard, and Laurent Daudet,
\newblock ``Late reverberation synthesis: From radiance transfer to feedback
  delay networks,''
\newblock {\em IEEE Trans. Audio, Speech, Lang. Process.}, vol. 23, no. 12, pp.
  2260--2271, 2015.

\bibitem{aknin2021stochastic}
Achille Aknin and Roland Badeau,
\newblock ``Stochastic reverberation model with a frequency dependent
  attenuation,''
\newblock in {\em Proc. WASPAA}. IEEE, 2021, pp. 351--355.

\bibitem{lalay_modephysique_2025}
Louis Lalay, Mathieu Fontaine, and Roland Badeau,
\newblock ``Modèle physique variationnel pour l'estimation de réponses
  impulsionnelles de salles (in french),''
\newblock in {\em 30ème Colloque sur le traitement du signal et des images},
  Strasbourg, August 25 - August 29 2025, number 2025-1533, pp. p. 393--396,
  GRETSI - Groupe de Recherche en Traitement du Signal et des Images.

\bibitem{belhomme}
Arthur Belhomme, Roland Badeau, Yves Grenier, and Eric Humbert,
\newblock ``Amplitude and phase dereverberation of harmonic signals,''
\newblock in {\em Proc. WASPAA}, 2017, pp. 294--298.

\bibitem{tflocoformer}
Kohei Saijo, Gordon Wichern, Fran{\c{c}}ois~G. Germain, Zexu Pan, and Jonathan
  Le~Roux,
\newblock ``{TF}-locoformer: Transformer with local modeling by convolution for
  speech separation and enhancement,''
\newblock in {\em Proc. IWAENC}, 2024, pp. 205--209.

\bibitem{fullsubnet}
Xiang Hao, Xiangdong Su, Radu Horaud, and Xiaofei Li,
\newblock ``Fullsubnet: A full-band and sub-band fusion model for real-time
  single-channel speech enhancement,''
\newblock in {\em Proc. ICASSP}. June 2021, IEEE.

\bibitem{bahrman_ICASSP}
Louis Bahrman, Mathieu Fontaine, and Gaël Richard,
\newblock ``A hybrid model for weakly-supervised speech dereverberation,''
\newblock in {\em Proc. ICASSP}. IEEE, 2025, pp. 1--5.

\bibitem{importance-phase}
Kuldip Paliwal, Kamil Wójcicki, and Benjamin Shannon,
\newblock ``The importance of phase in speech enhancement,''
\newblock {\em Speech Communication}, vol. 53, no. 4, pp. 465--494, 2011.

\bibitem{gerkmann2015phase-processing}
Timo Gerkmann, Martin Krawczyk-Becker, and Jonathan Le~Roux,
\newblock ``Phase processing for single-channel speech enhancement: History and
  recent advances,''
\newblock {\em IEEE Signal Process. Mag.}, vol. 32, no. 2, pp. 55--66, 2015.

\bibitem{tf-gridnet}
Zhong-Qiu Wang, Samuele Cornell, Shukjae Choi, Younglo Lee, Byeong-Yeol Kim,
  and Shinji Watanabe,
\newblock ``{TF-GRIDNET}: Making time-frequency domain models great again for
  monaural speaker separation,''
\newblock in {\em Proc. ICASSP}. IEEE, 2023, pp. 1--5.

\bibitem{schwartz-mag-or-phase}
Ayal Schwartz, Sharon Gannot, and Shlomo~E. Chazan,
\newblock ``Magnitude or phase? {A} two-stage algorithm for single-microphone
  speech dereverberation,''
\newblock in {\em Proc. IWAENC}, 2024, pp. 454--458.

\bibitem{griffin-lim}
Daniel~W. Griffin and Jae Lim,
\newblock ``Signal estimation from modified short-time {F}ourier transform,''
\newblock {\em IEEE Trans. Acoust., Speech, Signal Process.}, vol. 32, no. 2,
  pp. 236--243, 1984.

\bibitem{skerry-ryan2017tacotron}
RJ~Skerry-Ryan, Daisy Stanton, Yonghui Wu, Ron Weiss, Navdeep Jaitly, Zongheng
  Yang, Ying Xiao, Zhifeng Chen, Samy Bengio, Quoc Le, Yannis Agiomyrgiannakis,
  Rob Clark, and Rif Saurous,
\newblock ``Tacotron: Towards end-to-end speech synthesis,''
\newblock in {\em Proc. Interspeech}, August 2017, pp. 4006--4010.

\bibitem{li2021two-heads-are-better-than-one}
Andong Li, Wenzhe Liu, Chengshi Zheng, Cunhang Fan, and Xiaodong Li,
\newblock ``Two heads are better than one: A two-stage complex spectral mapping
  approach for monaural speech enhancement,''
\newblock {\em IEEE Trans. Audio, Speech, Lang. Process.}, vol. 29, pp.
  1829--1843, 2021.

\bibitem{zhao2019two-stage-speech-enhancement}
Yan Zhao, Zhong-Qiu Wang, and DeLiang Wang,
\newblock ``Two-stage deep learning for noisy-reverberant speech enhancement,''
\newblock {\em IEEE Trans. Audio, Speech, Lang. Process.}, vol. 27, no. 1, pp.
  53--62, 2019.

\bibitem{ephraim1984speechMMSE}
Yariv Ephraim and David Malah,
\newblock ``Speech enhancement using a minimum-mean square error short-time
  spectral amplitude estimator,''
\newblock {\em IEEE Trans. Acoust., Speech, Signal Process.}, vol. 32, no. 6,
  pp. 1109--1121, 1984.

\bibitem{richter2024ears}
Julius Richter, Yi-Chiao Wu, Steven Krenn, Simon Welker, Bunlong Lay, Shinjii
  Watanabe, Alexander Richard, and Timo Gerkmann,
\newblock ``{EARS}: An anechoic fullband speech dataset benchmarked for speech
  enhancement and dereverberation,''
\newblock in {\em Proc. Interspeech}, 2024, pp. 4873--4877.

\bibitem{ARNI}
Karolina Prawda, Sebastian Schlecht, and Vesa Välimäki,
\newblock ``Robust selection of clean swept-sine measurements in non-stationary
  noise,''
\newblock {\em J. Acoust. Soc. Am.}, vol. 151, pp. 2117--2126, March 2022.

\bibitem{ACE-Challenge}
James Eaton, Nikolay~D. Gaubitch, Alastair~H. Moore, and Patrick~A. Naylor,
\newblock ``Estimation of room acoustic parameters: The {ACE} challenge,''
\newblock {\em IEEE Trans. Audio, Speech, Lang. Process.}, vol. 24, no. 10, pp.
  1681--1693, 2016.

\bibitem{AIR}
Marco Jeub, Magnus Schafer, and Peter Vary,
\newblock ``A binaural room impulse response database for the evaluation of
  dereverberation algorithms,''
\newblock in {\em 2009 16th International Conference on Digital Signal
  Processing}, 2009, pp. 1--5.

\bibitem{BRUDEX}
Daniel Fejgin, Wiebke Middelberg, and Simon Doclo,
\newblock ``Brudex database: Binaural room impulse responses with uniformly
  distributed external microphones,''
\newblock in {\em Speech Communication; 15th ITG Conference}, 2023, pp.
  126--130.

\bibitem{dEchorate}
Diego~Di Carlo, Pinchas Tandeitnik, Cedric Foy, Nancy Bertin, Antoine
  Deleforge, and Sharon Gannot,
\newblock ``d{E}chorate: a calibrated room impulse response dataset for
  echo-aware signal processing,''
\newblock {\em EURASIP J. on Audio, Speech, and Music Proc.}, vol. 2021, no. 1,
  pp. 39, 2021.

\bibitem{SISDR}
Jonathan Le~Roux, Scott Wisdom, Hakan Erdogan, and John~R Hershey,
\newblock ``{SDR}--half-baked or well done?,''
\newblock in {\em Proc. ICASSP}. IEEE, 2019, pp. 626--630.

\bibitem{SRMR}
Tiago~H Falk, Chenxi Zheng, and Wai-Yip Chan,
\newblock ``A non-intrusive quality and intelligibility measure of reverberant
  and dereverberated speech,''
\newblock {\em IEEE Trans. Audio, Speech, Lang. Process.}, vol. 18, no. 7, pp.
  1766--1774, 2010.

\bibitem{PESQ}
ITUT Rec,
\newblock ``Wideband extension to recommendation p. 862 for the assessment of
  wideband telephone networks and speech codecs,''
\newblock {\em ITU-T recommendation P}, vol. 862, 2005.

\bibitem{STOI}
Jesper Jensen and Cees~H Taal,
\newblock ``An algorithm for predicting the intelligibility of speech masked by
  modulated noise maskers,''
\newblock {\em IEEE Trans. Audio, Speech, Lang. Process.}, vol. 24, no. 11, pp.
  2009--2022, 2016.

\end{thebibliography}

\end{document}